\documentclass[showpacs,aps,prb,twocolumn]{revtex4-1}
\usepackage{amsfonts,hyperref}
\usepackage{graphicx}
\usepackage{amssymb}
\usepackage{epsfig}
\usepackage{amsmath}

\begin{document}

\title{Quantized Anomalous Hall Insulator in a Nanopatterned Two-Dimensional
Electron Gas}
\author{Yongping Zhang}
\author{Chuanwei Zhang}
\thanks{cwzhang@wsu.edu}

\begin{abstract}
We propose that a quantum anomalous Hall insulator (QAHI) can be realized in
a nanopatterned two-dimensional electron gas (2DEG) with a small in-plane
magnetic field and a high carrier density. The Berry curvatures originating
from the in-plane magnetic field and Rashba and Dresselhaus spin-orbit
coupling, in combination with a nanoscale honeycomb lattice potential
modulation, lead to topologically nontrivial insulating states in the 2DEG
without Landau levels. In the bulk insulating gaps, the anomalous Hall
conductivity is quantized $-e^{2}/h$, corresponding to a finite Chern number
$-1$. There exists one gapless chiral edge state on each edge of a finite
size 2DEG.

\end{abstract}
\affiliation{Department of Physics and Astronomy, Washington State
University, Pullman, WA, 99164 USA}
\pacs{73.43.-f, 73.21.-b, 72.25.Dc, 72.20.-i}
\maketitle


\section{Introduction}

The quantum Hall effect was observed in a two-dimensional electron gas in
the presence of a high perpendicular orbital magnetic field, which breaks
the time reversal symmetry (TRS) and leads to Landau levels for quantum
states. The TRS can also be broken spontaneously (\textit{e.g.}, in a
ferromagnetic phase) or by other means without the external orbital magnetic
field and the associated Landau levels. The TRS breaking without Landau
levels, together with spin-orbit coupling, yield the experimentally observed
anomalous Hall effects (AHE) in ferromagnetic semiconductors \cite%
{Nagaosa,Xiao}. The quantized version of the AHE, \textit{i.e.}, a quantum
anomalous Hall insulator (QAHI), is a band insulator with quantized charge
Hall conductivity but without Landau levels. The QAHI was first proposed by
Haldane in a honeycomb lattice with periodic magnetic fields that induce
circulating current loops within one unit cell \cite{Haldane,Onoda}.
Motivated by the study of quantum spin Hall insulator \cite%
{Bernevig,Kane,Fu1,Qi,Moore,Hasan,Qi2}, the QAHI has been predicted recently
to exist in HgMnTe quantum wells \cite{Liu}, BiTe topological insulators
\cite{Yu}, and graphene with Rashba spin-orbit coupling \cite{Qiao}. In
these proposals, the TRS is broken by an exchange Zeeman field induced by
uniformly doping or adsorbing transition metal atoms. Schemes for realizing
a QAHI using ultracold atoms in optical lattices \cite{Wu} were also
proposed. However, no experimental observation of the QAHI has been reported
hitherto.

In this paper, we propose to realize a QAHI in a nanopatterned 2DEG with a
high carrier density ($\sim 10^{11}$ cm$^{-2}$) and a small in-plane
magnetic field ($\sim $ 0.5 T). The electrons in the 2DEG are modulated by
an external nanoscale honeycomb lattice potential \cite{Park}, which has
been experimentally realized in many solid state systems \cite%
{Gibertini,Simoni,Chou,Dorp}. In our scheme, the required TRS breaking for
the QAHI is achieved using a small in-plane magnetic field, which produces
an in-plane Zeeman field for electrons. Note that an in-plane magnetic field
does not induce orbital physics and the associated Landau levels. The 2DEG
is confined in a semiconductor quantum well grown along the specific (110)
direction \cite{Ohno,Sih,Alicea} with both Rashba and Dresselhaus spin-orbit
coupling. We show that a combination of the honeycomb lattice modulation,
Rashba and Dresselhaus spin-orbit coupling, and the small in-plane magnetic
field, can open two topologically nontrivial bulk energy gaps in the
spectrum, in which the charge Hall conductivity is quantized $-e^{2}/h$
(corresponds to a finite Chern number $-1$). Here $e$ is the charge of
electron and $h$ is the Plank constant. We calculate the Berry curvature
distribution and characterize the dependence of the anomalous Hall
conductivity on the chemical potential. We show that there exist one gapless
chiral edge state on each edge of a finite size 2DEG, consisting with the
Chern number $-1$. Finally, we show that the QAHI can be observed in a wide
range of experimentally feasible parameters.

There are two main advantages of using the nanopatterned 2DEG for realizing
a QAHI: 1) The 2DEG has been studied for several decades and accumulated
technology for semiconductors makes it a promising experimental candidate.
One outstanding feature of a 2DEG is its high purity, leading to high
mobility of electrons. 2) The small in-plane magnetic field ($\sim $ 0.5T)
can be generated using permanent bar magnets or magnetic current coils,
which are much easier to implement and tune to a large magnitude in
experiments than doping or adsorbing transition metal elements in
topological insulator or graphene \cite{Liu,Yu,Qiao}. The bar magnets or
coils can be easily integrated into the design of the QAHI, and their great
tunability (compared to doping magnetic atoms) may lead to new functionality
for device applications.

The paper is organized as follows: Section II illustrates the proposed
experimental setup for the observation of QAHI. Section III describes the
QAHI in a 2DEG in the presence of a honeycomb lattice potential modulation
and an in-plane magnetic field. The experimental parameters for the
observation of the QAHI is also discussed. Section IV is the conclusion.

\section{Proposed experimental scheme}

\begin{figure}[t]
\includegraphics[width=1\linewidth]{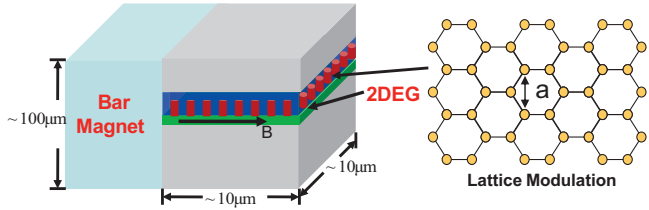} \vspace{-10pt}
\caption{(Color online) Schematic illustration of the proposed experimental
scheme. The 2DEG confined in a quantum well grown along (110) direction,
which possesses Rashba and Dresselhaus spin-orbit coupling simultaneously. A
small in-plane magnetic field, generated by bar magnets or magnetic current
coils, is applied to generate an in-plane Zeeman field. There are nanoscale
periodic modulation pillars (radius $b$) with a honeycomb lattice structure
on top of the 2DEG.}
\label{setup}
\end{figure}

The proposed experimental scheme is illustrated in Fig. \ref{setup}. A 2DEG
is confined in a quantum well grown along the (110) direction \cite{Ohno,Sih}
with the layer inversion symmetry explicitly broken by imbalancing the
quantum well using a gate voltage and/or chemical means. The Rashba and
Dresselhaus spin-orbit coupling coexist in such a 2DEG, and the dynamics of
electrons in the presence of an in-plane magnetic field can be described by
the Hamiltonian \cite{Alicea},
\begin{equation}
H_{0}=\frac{\hbar ^{2}k^{2}}{2m^{\ast }}+\alpha _{R}(k_{y}\sigma
_{x}-k_{x}\sigma _{y})+\alpha _{D}k_{x}\sigma _{z}+h_{y}\sigma _{y},
\label{Hamiltonian1}
\end{equation}%
where $m^{\ast }$ is the effective mass of electrons in the conduction band,
$\alpha _{R}$ and $\alpha _{D}$ are the strengths of Rashba and Dresselhaus
spin-orbit coupling respectively, $\hat{\sigma}=(\sigma _{x},\sigma
_{y},\sigma _{z})$ are Pauli matrices, and $h_{y}$ is the magnitude of the
in-plane Zeeman field induced by a small in-pane magnetic field that can be
generated by bar magnets or magnetic current coils. The Hamiltonian (\ref%
{Hamiltonian1}) has two eigenenergies $E_{\pm \mathbf{k}}=\frac{\hbar
^{2}k^{2}}{2m^{\ast }}\pm E_{0}$ with
\begin{equation}
E_{0}=\sqrt{\alpha _{D}^{2}k_{x}^{2}+\alpha _{R}^{2}k_{y}^{2}+(\alpha
_{R}k_{x}-h_{y})^{2}}.
\end{equation}%
The minimum energy gap between two bands is $\Delta =2\alpha _{D}h_{y}/\sqrt{%
\alpha _{D}^{2}+\alpha _{R}^{2}}$ that occurs at $k_{x}=\alpha
_{R}h_{y}/(\alpha _{D}^{2}+\alpha _{R}^{2})$ and $k_{y}=0$. The gap is
nonzero only for nonzero $\alpha _{D}$ and $h_{y}$. The 2DEG is modulated by
a two-dimensional honeycomb lattice potential (Fig. \ref{setup})%
\begin{equation}
V\left( \mathbf{r}\right) =V_{0}\text{ if }\left\vert \mathbf{r}-\mathbf{r}%
_{j}\right\vert \leq b\text{; otherwise 0,}  \label{lattice}
\end{equation}%
where $\mathbf{r}_{j}$ is the honeycomb lattice point, $b$ is the radius of
the nanoscale pillar, and the lattice constant $a_{0}=a/\sqrt{3}$. The
lattice structure has been realized in a GaAs 2DEG in a recent experiment
\cite{Gibertini,Simoni}.

\section{QAHI in a 2DEG}

\subsection{2DEG without lattice modulation}

Before presenting the QAHI physics with the lattice modulation, we first
discuss the intrinsic AHE in the 2DEG without the lattice potential, which
originate from the nontrivial topological properties induced by the
spin-orbit coupling and the in-plane Zeeman field. The Rashba and
Dresselhaus spin-orbit coupling, together with the TRS breaking induced by
the in-plane Zeeman field, yield non-zero Berry curvatures $\mathbf{\Omega }%
_{\pm }=-2$Im$\langle \frac{\partial \Phi _{\pm }}{\partial k_{x}}|\frac{%
\partial \Phi _{\pm }}{\partial k_{y}}\rangle \mathbf{e}_{z}$ for electrons
in the two energy bands $E_{\pm \mathbf{k}}$. Here $\Phi _{\pm }$ are the
eigenfunctions of the Hamiltonian (\ref{Hamiltonian1}).

Straightforward calculation shows
\begin{equation}
\mathbf{\Omega }_{\pm }=\mp \alpha _{R}\alpha _{D}h_{y}\mathbf{e}%
_{z}/2E_{0}^{3},  \label{Berry}
\end{equation}%
which have the same amplitude but opposite sign in two bands. $\mathbf{%
\Omega }_{\pm }$ are equivalent to an effective magnetic field in momentum
space and yield an anomalous velocity of electrons $\mathbf{v}=e\mathbf{E}%
\times \mathbf{\Omega }_{\pm }$ when an electric field $\mathbf{E}$ is
applied \cite{Xiao}. The intrinsic AHE is the manifestation of the anomalous
velocity of electrons with the Hall conductivity%
\begin{equation}
\sigma _{H}=\frac{e^{2}}{h}\sum\nolimits_{n}\int \frac{dk_{x}dk_{y}}{2\pi }%
\Omega _{n}f(E_{n\mathbf{k}}),  \label{AHC}
\end{equation}%
where $f(E_{n\mathbf{k}})=1/(e^{(E_{n\mathbf{k}}-\mu )/k_{B}T}+1)$ is the
Fermi-Dirac distribution, $\mu $ the chemical potential, $n=\pm $ is the
band index, $k_{B}$ is the Boltzmann constant and $T$ is the temperature.

\begin{figure}[t]
\includegraphics[width=1\linewidth]{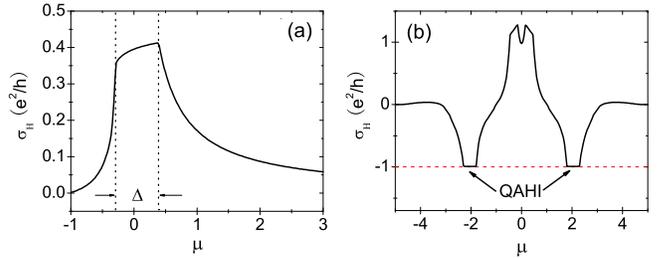} \vspace{-20pt}
\caption{Plot of the anomalous charge Hall conductivity $\protect\sigma _{H}$
versus the chemical potential $\protect\mu $ without (a) or with (b) lattice
modulation. (a) $\protect\alpha _{D}k_{c}=0.8,$ $\protect\alpha %
_{R}k_{c}=0.8,$ $h_{y}=0.5$. The space between dashed lines is the gap $%
\Delta $ between two energy bands. The energy unit is chosen as $\hbar
^{2}k_{c}^{2}/m^{\ast }$, with $k_{c}=3$ nm$^{-1}$ for $\protect\alpha %
_{D}=0.3$ eV$\cdot $A. (b) $\protect\lambda _{R}=0.3,\protect\lambda %
_{D}=0.8 $, $h_{y}=1.6$. The energy unit is tunneling amplitude $t$.}
\label{HallConductivity}
\end{figure}

Fig. \ref{HallConductivity}a shows the dependence of $\sigma _{H}$ on the
chemical potential $\mu $ at zero temperature. $\mu $ can be tuned in
experiments by varying the gate voltage of the quantum well. When $\mu $
lies just above the energy minimum of the lower band, the Berry curvature
for occupied electrons is small and $\sigma _{H}$ is small. When $\mu $
sweeps across the energy gap between two bands, $\sigma _{H}$ reaches the
maximum. Further increase of $\mu $ leads to the occupation of electrons in
the upper band and the decrease of $\sigma _{H}$ due to the cancellation of
the contributions from two bands with opposite sign of the Berry curvatures.
When $\mu \rightarrow \infty $, the total $\sigma _{H}$ approaches zero.

\subsection{QAHI with lattice modulation}

A QAHI is a band insulator with a bulk energy gap opened in the spectrum.
Generally, a bulk energy gap can be obtained by coupling electrons to a
periodic lattice potential, which transforms the plane wave states to the
Bloch states of electrons. In this Letter, we consider a honeycomb lattice
potential modulation (\ref{lattice}) of the 2DEG in the tight-binding
region, where the free space Hamiltonian (\ref{Hamiltonian1}) changes to
\begin{align}
H=& -t\sum_{\langle ij\rangle }c_{i}^{\dag }c_{j}+i\lambda _{R}\sum_{\langle
ij\rangle }c_{i}^{\dag }\left[ \mathbf{\hat{\sigma}}\times \mathbf{d}%
_{ij}\cdot \mathbf{e}_{z}\right] c_{j}  \notag \\
& +i\lambda _{D}\sum_{\langle ij\rangle }c_{i}^{\dag }\left[ \mathbf{\hat{%
\sigma}}\times \mathbf{d}_{ij}\cdot \mathbf{e}_{y}\right] c_{j}+h_{y}%
\sum_{i}c_{i}^{\dag }\sigma _{y}c_{i},  \label{TBH}
\end{align}%
where $c_{i}^{\dag }(c_{i})$ creates (annihilates) an electron at site $i$, $%
\langle ij\rangle $ represents two nearest neighboring sites, $t=\int d^{2}%
\mathbf{r\Phi }^{\ast }\left( \mathbf{r}\right) V\left( \mathbf{r}\right)
\mathbf{\Phi }\left( \mathbf{r-a}\right) $ is the spin-independent nearest
neighbor tunneling strength, $\mathbf{\Phi }\left( \mathbf{r}\right) $ is
the eigenstates of a single pillar centered at the origin and $\mathbf{a}$
is a nearest-neighbor position vector. For $b/a_{0}\sim 0.23$, $t/E_{L}\sim
0.06\exp \left( -V_{0}/0.85E_{L}\right) $ \cite{Gibertini,Simoni}, where $%
E_{L}=\frac{\hbar ^{2}}{2m^{\ast }}\left( \frac{2\pi }{a_{0}}\right) ^{2}$
is the energy defined by the lattice constant $a_{0}$. $\lambda _{R}$ and $%
\lambda _{D}$ are the spin-dependent nearest neighbor tunneling strengths
originating from Rashba and Dresselhaus spin-orbit coupling, $\mathbf{d}%
_{ij} $ is the unit bond vector between nearest neighboring sites $i$ and $j$%
, $h_{y}$ is the in-plane Zeeman field. Henceforth we set $t=1$ as the
energy unit. \
\begin{figure}[t]
\includegraphics[width=1.0\linewidth]{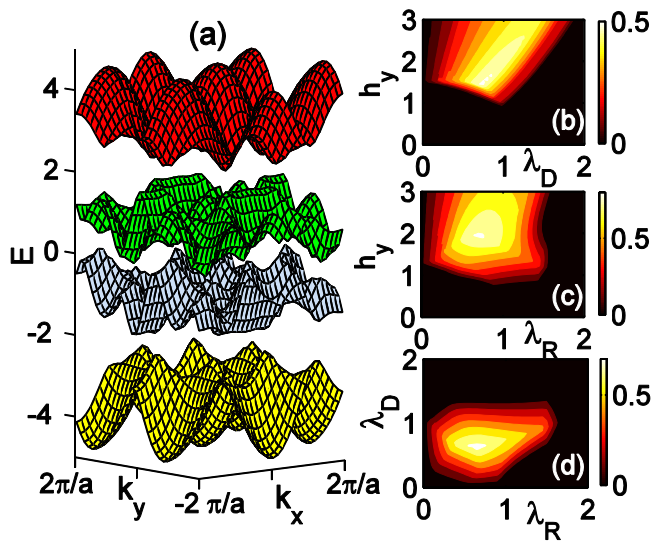} \vspace{-20pt}
\caption{(Color online) (a) Plot of the energy spectrum of the tight-binding
Hamiltonian (\protect\ref{TBH}). $\protect\lambda _{R}=0.3$, $\protect%
\lambda _{D}=0.8$ and $h_{y}=1.6$. (b)-(d) Plot of the bulk energy gap
between the lowest two bands for different parameters. (b) $\protect\lambda %
_{R}=0.3$. (c) $\protect\lambda _{D}=0.8$. (d) $h_{y}=1.5$.}
\label{gap}
\end{figure}

The energy spectrum is obtained by numerically diagonalizing the
tight-binding Hamiltonian (\ref{TBH}) in momentum space through the Fourier
transformation $c_{i}\left( \tau ,\sigma \right) =\sum_{\mathbf{k}}c_{%
\mathbf{k}}\left( \tau ,\sigma \right) \exp \left( -i\mathbf{k}\cdot \mathbf{%
R}_{i}\right) $, assuming a periodic boundary condition of the system. Here $%
\tau =\{A,B\}$ represent two sublattice degrees of freedom of electrons in a
honeycomb lattice, and $\sigma =\{\uparrow ,\downarrow \}$ are the spin
degrees of freedom. These four degrees of freedom lead to four energy bands
of the Hamiltonian (\ref{TBH}), as plotted in Fig. \ref{gap}a. There are
energy gaps opened between the lowest two bands and the highest two bands
respectively. The gaps exist for a wide range of physical parameters. Taking
account of the symmetry between two gaps, we only need characterize the gap
between the lowest two bands, which is plotted in Figs. \ref{gap}b-\ref{gap}%
d for different parameters. Figs. \ref{gap}b and \ref{gap}c have constant $%
\lambda _{R}=0.3$ and $\lambda _{D}=0.8$ respectively. Clearly, the gap is
opened only above a critical value of $h_{y}$. Fig. \ref{gap}d is plotted
with a fixed $h_{y}=1.5$. The gap only exists in certain area in the
parameter plane spanned by $\lambda _{R}$ and $\lambda _{D}$. The presence
of the bulk energy gap indicates an insulating state of the system. Another
interesting feature of our system is that there is no gap between the second
and third bands. Instead, the valley centers at these two bands shift to
opposite directions due to the lack of in-plane rotation symmetry of the
Hamiltonian (\ref{TBH}). Although these two bands do not touch at any $%
\mathbf{k}$, a full energy gap does not exist. This is very different from
the graphene scheme with Rashba spin-orbit coupling and perpendicular
exchange field, where the gap is opened between the second and third bands
\cite{Qiao}.

\begin{figure}[b]
\includegraphics[width=0.6\linewidth]{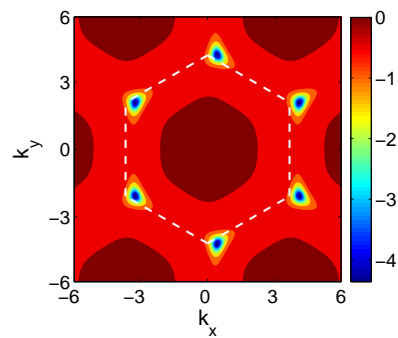} \vspace{-10pt}
\caption{(Color online) Contour plot of the Berry curvature in the lowest
band. $\protect\lambda _{R}=0.3,\protect\lambda _{D}=0.8$, $h_{y}=1.6$. The
dashed line is the boundary of the first Brillouin zone of the honeycomb
lattice.}
\label{BerryC}
\end{figure}

The nanopatterned 2DEG inherits the topological properties of the 2DEG
without the lattice modulation. We numerically calculate the Berry curvature
distribution in the lowest band and plot it in Fig. \ref{BerryC}. The peaks
of the Berry curvatures shift toward the positive $k_{x}$ direction from the
corners of the Brillouin zone, which can be understood from the fact that
the Berry curvature (\ref{Berry}) without the lattice modulation has a peak
at $k_{x}=\alpha _{R}h_{y}/(\alpha _{D}^{2}+\alpha _{R}^{2})$ due to the
in-plane Zeeman field and Dresselhaus spin-orbit coupling that destroy the
rotation symmetry in the plane.

When the chemical potential lies inside the bulk gap, the 2DEG is a
insulator and there is no longitudinal current with an applied electric
field. However, as we have discussed in the case without lattice modulation,
the non-zero Berry curvature can yield a transversal Hall current by
inducing an anomalous velocity of electrons. The anomalous Hall conductivity
$\sigma _{H}$, calculated from Eq. (\ref{AHC}) by integrating the Berry
curvature over the first Brillouin zone, is found to be quantized $-e^{2}/h$
in both gaps (Fig. \ref{HallConductivity}b). The quantized $\sigma _{H}$ is
directly related to the topological invariant of the system known as the
first Chern number $\mathcal{C}$ \cite{Thouless}. The Hall conductivity can
be expressed as $\sigma _{H}=\frac{e^{2}}{h}\mathcal{C}$, with
\begin{equation}
\mathcal{C}=\int_{BZ}\frac{d^{2}k}{2\pi }\Omega
\end{equation}%
as the integral over the first Brillouin zone. Therefore $\mathcal{C}=-1$
for both gaps in the nanopatterned 2DEG. Note that here the quantized
anomalous Hall conductivity originates from the interplay between the
sublattice and spin degrees of freedom of electrons in the honeycomb
lattice. We have checked that a simple square lattice modulation does not
lead to the quantized $\sigma _{H}$.

\begin{figure}[t]
\includegraphics[width=0.6\linewidth]{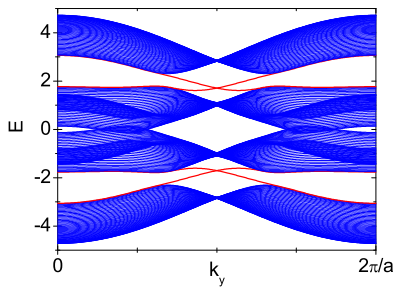} \vspace{-10pt}
\caption{(Color online) Band spectrum of a nanopatterned 2DEG with a zigzag
boundary along the $x$ direction. $\protect\lambda _{R}=0.3,\protect\lambda %
_{D}=0.8$, $h_{y}=1.6$. The red lines are the gapless chiral edge states.}
\label{Edge}
\end{figure}

In a QAHI, currents flowing along the boundary of the system form gapless
chiral edge states within the gap, which are the manifestation of the
quantized Hall conductivity obtained from the bulk topological property. The
emergence of gapless chiral edge states can be obtained from the energy
spectrum of a finite two dimensional system. We construct a finite size
nanopatterned 2DEG that has a zigzag boundary along the $x$ direction and is
infinite along the $y$ direction. The resulting energy spectrum is shown in
Fig. \ref{Edge}. There are still two gaps opened between the lowest two
bands and the highest two bands. Within each gap, there exist two gapless
chiral edge states on two edges of this finite size 2DEG. The number of edge
states on each edge is equal to the absolute value of the first Chern number
$\left\vert \mathcal{C}\right\vert =1$, as expected. The chiral edge
currents can be probed in a typical six-probe Hall-bar experimental setup.

\subsection{Experimental parameters}

In experiments, the honeycomb potential modulation on a 2DEG with a typical
lattice constant $a_{0}\sim 130$ nm has been realized \cite{Gibertini,Simoni}%
. More generally, nanostructures with the square lattice geometry and a
typical lattice constant $a_{0}\sim 10$ nm have also been realized in many
experiments \cite{Chou,Dorp}. Implementing honeycomb lattice structures with
similar small lattice constants should be straightforward. We consider InSb
semiconductor quantum wells with large spin-orbit coupling. The effective
mass of electrons in the conduction bands is $m^{\ast }=0.014m_{e}$,
therefore the corresponding energy unit of the lattice system is $E_{L}=%
\frac{\hbar ^{2}}{2m^{\ast }}\left( \frac{2\pi }{a_{0}}\right) ^{2}\sim 6.2$
meV for $a_{0}\sim 130$ nm, and $b\sim 30$ nm. For a lattice depth $%
V_{0}=0.6E_{L}\sim 3.7$ meV and electron density $n\sim 5\times 10^{9}$ cm$%
^{-2}$ (roughly one electron per site), the nanopatterned 2DEG is in the
tight-binding region, with a typical tunneling strength between neighboring
lattice sites $t\sim 0.03E_{L}\sim 0.2$ meV. For $a_{0}\sim 130$ nm, the
typical Rashba and Dresselhaus spin-orbit coupling strength that can be
reached in an InSb 2DEG are $\lambda _{R}\sim 0.08$ meV, and $\lambda
_{D}\sim 0.2$ meV \cite{Alicea} , which are within the parameter region
where the band gaps are opened. The InSb also has a large $g$ factor $g\sim
70$ \cite{gfactor}, which ensures an in-plane Zeeman energy $h_{y}\sim 0.25$
meV can be easily realized with a small magnetic field $\sim 0.1$ T
generated by bar magnets or magnetic current coils. The resulting gap size
is $\sim 0.15$ meV, which corresponds to $\sim 2$ K temperature and can be
easily achieved in a 2DEG.

The size of the gap and the density of electrons can be further increased by
using a smaller lattice constant in experiments. For instance, for $%
a_{0}\sim 30$ nm, $b\sim 7$ nm with corresponding spin-orbit coupling
energies $\lambda _{R}\sim 0.35$ meV, $\lambda _{D}\sim 0.8$ meV, a set of
physical parameters within current experimental technology are $E_{L}\sim
116 $ meV, $B\sim 0.5$ T, $h_{y}\sim 1$ meV, $n\sim 1.3\times 10^{11}$ cm$%
^{-2}$, and $t\sim 0.008E_{L}\sim 0.8$ meV for $V_{0}\sim 1.8E_{L}$. The
resulting gap is now $\sim 0.5$ meV ($\sim 6$ K). With such a high carrier
density $n$ and the high mobility of electrons in the 2DEG \cite{mobility},
the effects of disorder can be neglected \cite{Simoni,mobility}. To reduce
the required magnitude of $V_{0}$, a two-dimensional hole gas may be used,
which has larger effective mass $m_{h}^{\ast }=0.43m_{e}$, leading to much
smaller $E_{L}^{h}=0.03E_{L}$. Finally, we remark that although quantum Hall
effects have been observed in graphene at a medium magnetic field $\sim 2$ T
\cite{Jiang}, the Hall plateau is small and the corresponding Hall
conductivity $\sigma _{H}=14e^{2}/h$ (a very large magnetic field ($>10$ T)
is needed to reach $\sigma _{H}=e^{2}/h$).$\,$

\section{Conclusion}

In summary, we find that a QAHI can be realized in a 2DEG with a high
carrier density, a small magnetic field, and a nanoscale honeycomb lattice
potential modulation. We show that topologically nontrivial band energy gaps
can be opened, within a wide range of experimentally feasible parameters,
through a combination of the honeycomb lattice, Rashba and Dresselhaus
spin-orbit coupling, and a small in-plane magnetic field. Quantum anomalous
Hall effects can be observed in the bulk gaps with a quantized Hall
conductivity $\sigma _{H}=-e^{2}/h$, corresponding to a Chern number $C=-1$.

\textbf{Acknowledgement:} We thank Vito Scarola for valuable discussion.
This work is supported by ARO (W911NF-09-1-0248), DARPA-MTO
(FA9550-10-1-0497), and DARPA-YFA (N66001-10-1-4025). Y.Z. is also supported
by DOE (DE-FG02-02ER45958, Division of Materials Science and Engineering)
and Welch Foundation (F-1255).

\end{document}